\documentclass[twoside]{dis09}

\usepackage[latin1]{inputenc}

\usepackage[dvips]{graphicx,epsfig,color}

\usepackage{wrapfig,rotating}

\usepackage{amssymb,amsmath,array}

\usepackage{multirow}
\usepackage{url}

\newcommand{\tw}{\textwidth}
\newcommand{\be}{\begin{equation}\nonumber}
\newcommand{\ee}{\end{equation}}
\newcommand{\bea}{\begin{eqnarray}}
\newcommand{\eea}{\end{eqnarray}}

\newcommand{\la}{\left\langle}
\newcommand{\ra}{\right\rangle}
\newcommand{\lc}{\left[}
\newcommand{\rc}{\right]}
\newcommand{\lp}{\left(}
\newcommand{\rp}{\right)}

\newcommand{\bc}{\begin{center}}
\newcommand{\ec}{\end{center}}

\newcommand{\bi}{\begin{itemize}}
\newcommand{\ei}{\end{itemize}}

\newcommand{\rep}{\mathrm{rep}}

\definecolor{red}{rgb}{1,0,0}
\definecolor{green}{rgb}{0.3,0.6,0.3}
\definecolor{blue}{rgb}{0,0,1}
\definecolor{darkgreen}{rgb}{0,0.39,0.00}

\begin{document}

\title{The NNPDF1.2 parton set: implications for the LHC}




\author{Alberto~Guffanti~$^1$, Juan~Rojo$^2$ and Maria~Ubiali$^3$
%
%
\vspace{.3cm}\\
%
~$^1$  Physikalisches Institut, Albert-Ludwigs-Universit\"at Freiburg
\\ Hermann-Herder-Stra\ss e 3, D-79104 Freiburg i. B., Germany  \\
~$^2$ Dipartimento di Fisica, Universit\`a di Milano and
INFN, Sezione di Milano,\\ Via Celoria 16, I-20133 Milano, Italy\\
~$^3$ School of Physics and Astronomy, University of Edinburgh,\\
JCMB, KB, Mayfield Rd, Edinburgh EH9 3JZ, Scotland\\
}




\maketitle

\begin{abstract}


Recently a new set of Parton Distribution Functions (NNPDF1.2) has been produced and 
released by the NNPDF Collaboration.
The inclusion of dimuon data in the analysis allows a determination of the strange
content of the proton with faithful uncertainty estimation together with a precision determination 
of electroweak parameters.
In this contribution, we discuss some of the implications of the NNPDF1.2 set, and in 
particular of its uncertainty determination of the strange PDFs, for LHC phenomenology. 
First of all, we study the impact on the electroweak boson production cross-section,
with special attention to the $\sigma(Z)/\sigma(W)$ ratio.
Then we revisit the top pair production cross-section,
and perform a comparison of partonic fluxes between various
PDF sets. Finally,  we discuss the potential of using associated production of $W$ with a
charm quark at the Tevatron and the LHC to constrain the proton strangeness.
\end{abstract}

\section{The NNPDF1.2 parton set}

The determination of the strange and antistrange quark distributions
of the nucleon is considerably interesting from the phenomenological point of view.
However, till very recently, the bulk of data included in parton determinations, 
namely neutral-current deep-inelastic scattering, had minimal sensitivity to flavour 
separation, and no sensitivity at all to the separation of quarks and antiquark 
contributions. As a consequence, in standard parton fits the strange 
and antistrange quark distributions were not determined directly:
rather, they were assumed to be equal and proportional to the total light antiquark 
sea distribution.

Due to the availability of the new deep-inelastic neutrino and anti-neutrino 
charm production data, which is directly sensitive to the strange and antistrange
parton distributions, independent parametrizations of the strange and antistrange
distributions have been included in most recent parton fits. 

However, the standard method for determining parton distributions,
based on fitting the parameters of a fixed functional form, is known to be hard to 
handle when the experiments are relatively unconstraining. 
An alternative approach to parton determination which overcomes this
difficulty has been developed by the NNPDF Collaboration in a series
of papers~\cite{f2ns,f2p,DelDebbio:2007ee,Ball:2008by,Rojo:2008ke,nnpdf12}
\footnote{See also Ref.~\cite{Dittmar:2009ii} 
for a series of benchmark comparisons between the
NNPDF approach and the standard approach.}. 
The method is based on the use of neural networks for parton
parametrization, and a Monte Carlo method supplemented by a suitable training and
stopping algorithm for the construction of the parton fit. In this approach, 
parton distributions are given as a Monte Carlo sample which represents
their probability distributions as inferred from the data.

Recently, in Ref.~\cite{nnpdf12}, a new  parton set, NNPDF1.2, was 
constructed.\footnote{The NNPDF1.2 set are available in {\tt LHAPDF}
starting from version 5.7.1.}
The addition of dimuon data to the global inclusive deep-inelastic
 scattering dataset, on 
which the NNPDF1.0~\cite{Ball:2008by} parton set was based, allows
 a determination of the strange and 
antistrange distributions with faithful uncertainty estimation.

\begin{figure}[t]
  \centering
    \includegraphics[width=0.49\tw]{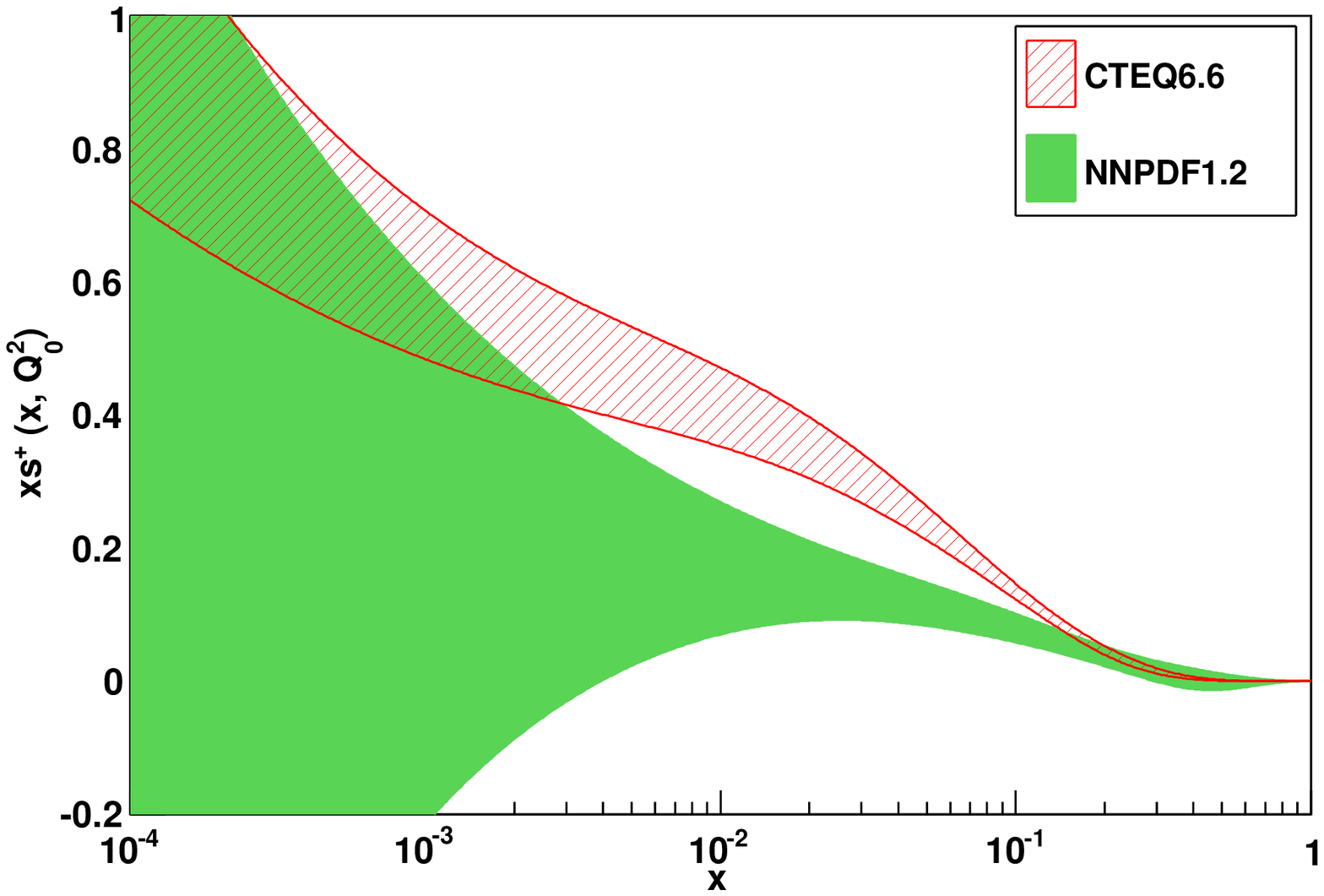}
    \includegraphics[width=0.49\tw]{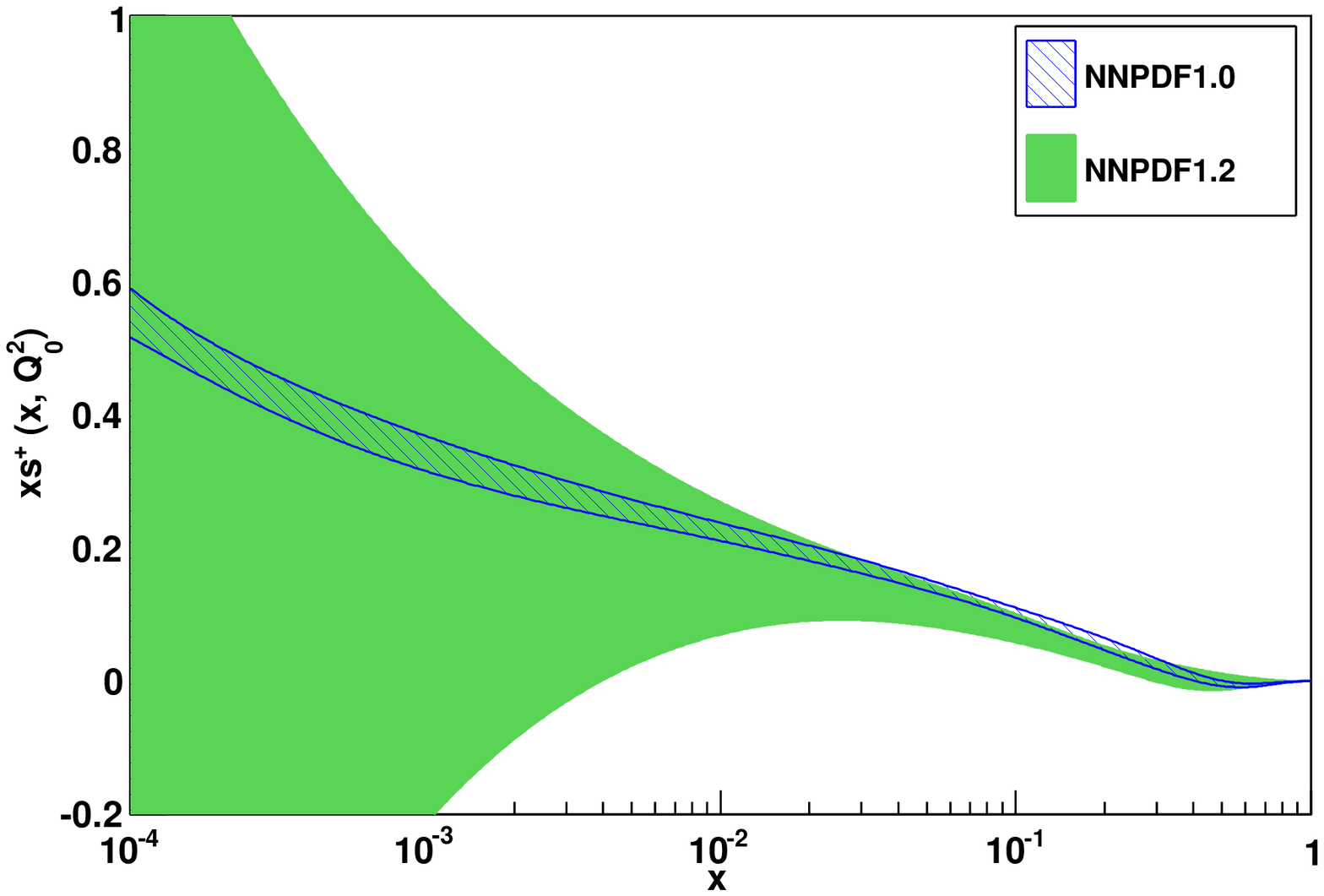}
    \includegraphics[width=0.49\tw]{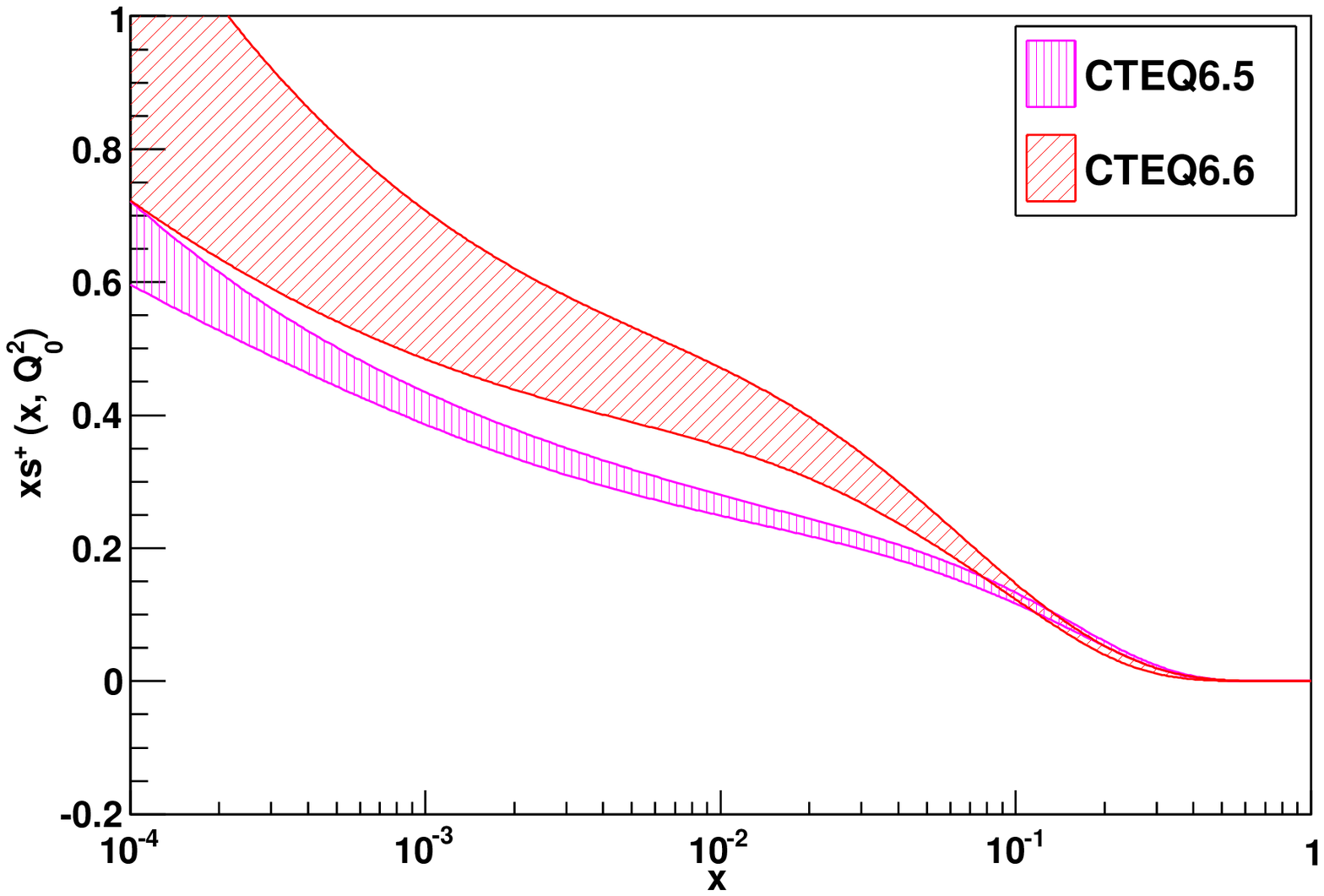}
    \includegraphics[width=0.49\tw]{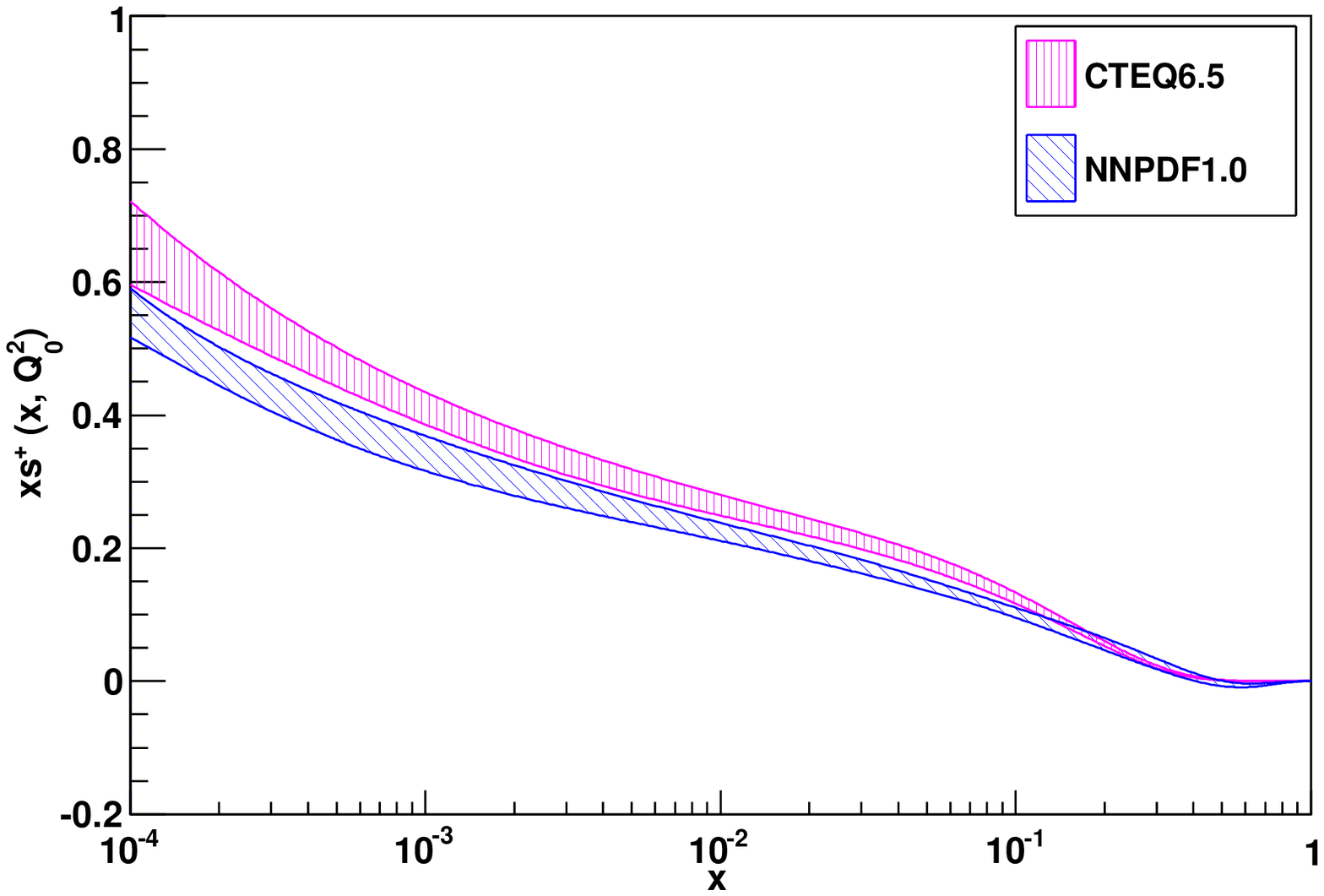}
  \caption{\small Comparison of the
 $s^+(x,Q_0^2)$ PDF from various analyses.
Upper left plot: NNPDF1.2 vs. CTEQ6.6, where in both cases
$s^+$ is determined from experimental data. Upper right plot:
NNPDF1.2, with fitted $s^+$  vs. NNPDF1.0, with $s^+$ fixed by
flavour assumptions.  Lower left plot:
CTEQ6.6, with fitted $s^+$  vs. CTEQ6.5, with $s^+$ fixed by
flavour assumptions. Lower right plot: 
NNPDF1.0 vs. CTEQ6.5, where in both cases
$s^+$ is fixed by
flavour assumptions. }
  \label{fig:s+}
\end{figure}

In Fig.~\ref{fig:s+} we compare the total strange PDF, $s^+(x,Q_0^2)$,
with $Q_0^2=$ 2 GeV$^2$, as determined from the NNPDF1.2 analysis with the results 
obtained from other parton sets: CTEQ6.6~\cite{Nadolsky:2008zw}, 
where an independent parametrization for $s^+$ is also used, and NNPDF1.0
and CTEQ6.5~\cite{Tung:2006tb}, where the strange PDFs are
fixed to be proportional to the non-strange sea distributions.
We observe that in the case of the NNPDF analysis the main effect of
releasing the strange parametrization is a substantial enlargement of 
the error on the $s^+$ distribution obtained in the NNPDF 1.2
analysis, which remains anyway compatible with the one of the NNPDF
1.0 set.
This is not the case for the CTEQ analysis where it can be
observed that the $s^+$ distributions from CTEQ 6.6 and CTEQ 6.5 present substantial
differences both in normalizations and in shapes, especially in the intermediate 
and small-$x$ region.

We refer to Ref.~\cite{nnpdf12} for further details on the NNPDF1.2
parton set and for a detailed description of its implications in the
precision determination of electroweak parameters, such as the CKM
matrix elements $|V_{cs}|$ and $|V_{cd}|$ and the QCD corrections 
to the Paschos-Wolfenstein relation.

Our purpose in this contribution is to briefly examine some of the 
phenomenological implications of the NNPDF1.2 parton set for LHC
physics. In particular, we assess the impact on standard candles 
of the increased uncertainty in the strange PDF as compared to
previous determinations. 

When comparing results computed using NNPDF sets with the ones obtained
using other parton sets, like CTEQ 6.5/6.6, it is important to keep in
mind that central values are affected by a systematic uncertainty due
to an approximate treatment of heavy flavour contributions. On the other
hand, the size of the PDF
uncertainties on the same predictions should not
be affected by the heavy flavour contribution.

\section{Impact of strangeness on LHC standard candles}

We begin by comparing results for the total cross-sections for weak
bosons production at the LHC ($\sigma\lp W^{\pm}\rp$ and $\sigma\lp
Z\rp$), obtained using the NNPDF 1.2 set with the ones 
obtained using CTEQ6.6,
NNPDF1.0, CTEQ6.5 and CTEQ6.1~\cite{Pumplin:2002vw} sets. 
The computation of the cross sections has been performed including NLO 
QCD corrections using the MCFM\footnote{Note that our
results for CTEQ 6.6 do not coincide with the results presented in 
Ref.~\cite{Nadolsky:2008zw} because of the differences in the codes 
used for the computation of physical observables.} program.
Results are collected in Table~\ref{tab:LHCobs}.

It is interesting to notice that the size of the PDF uncertainties on these
observables is very similar among all the different parton sets. This 
reinforces the observation already done in  Ref.~\cite{Nadolsky:2008zw}
that the uncertainty on total cross sections for weak boson production
is rather insensitive to the uncertainty on the strange distribution.

\begin{table}
  \centering
  {\scriptsize
  \begin{tabular}{|c|c|c|c|}
    \hline
         & $\sigma(W^+){\rm Br}\lp W^+ \to l^+\nu_l\rp\,$ [nb]
         & $\sigma(W^-){\rm Br}\lp W^- \to l^+\nu_l\rp\,$ [nb]
         & $\sigma(Z^0){\rm Br}\lp Z^0 \to l^+l^-\rp\,$ [nb] \\
    \hline
\hline
    \multirow{1}{*}{NNPDF 1.0} & 11.83 $\pm$ 0.26 & 8.41 $\pm$ 0.20 & 1.95 $\pm$ 0.04 \\
    \hline
    \multirow{1}{*}{NNPDF 1.2} & 11.99 $\pm$ 0.34 & 8.47 $\pm$ 0.21 & 1.97 $\pm$ 0.04 \\
    \hline
    \hline
    \multirow{1}{*}{CTEQ6.1}   & 11.65 $\pm$ 0.34 & 8.56 $\pm$ 0.26 & 1.93 $\pm$ 0.06 \\
    \hline
    \multirow{1}{*}{CTEQ6.5}   & 12.54 $\pm$ 0.29 & 9.19 $\pm$ 0.22 & 2.08 $\pm$ 0.04 \\
    \hline
    \multirow{1}{*}{CTEQ6.6}   & 12.41 $\pm$ 0.28 & 9.11 $\pm$ 0.22 & 2.07 $\pm$ 0.05 \\
\hline
  \end{tabular}}
  \caption{\small The LHC benchmark cross sections for  NNPDF1.0 and
    CTEQ6.5 (with strangeness proportional to the non-strange sea) 
    and NNPDF1.2 and CTEQ6.6 (with strangeness determined from the
    global analysis).
    The CTEQ6.1, with a ZM scheme for heavy quarks, is also shown for
    comparison. All numbers shown are for $\sqrt{s}$=14 TeV. PDF
    uncertainties correspond to 68\% confidence levels.
\label{tab:LHCobs}}
\end{table}

We turn then to a study of the sensitivity of the $Z/W$ ratio $r_{ZW}\equiv
\sigma_{Z}/(\sigma_{W^+}+\sigma_{W^-})$ to the uncertainty in the
strange distribution. This observable is particularly interesting
given that PDF uncertainties are greatly reduced when considering the
ratio of two cross-sections, and thus  provides an excellent candidate for a
measurement of the LHC luminosity. However, in
Ref.~\cite{Nadolsky:2008zw} it was shown that, although the impact of
the uncertainties on the $W$ and $Z$ cross sections due to the strange
PDF are rather small on their own, the ratio $r_{ZW}$ has a greater
sensitivity to the strange uncertainty in the region $0.01 < x < 0.05$
because PDF uncertainties do not completely cancel in the ratio. 
Therefore, $r_{ZW}$ is potentially affected by the larger uncertainty 
of strange PDF found in the NNPDF1.2 analysis with respect to other 
parton densities determinations.
 
The NNPDF 1.2 prediction for the ratio $r_{ZW}$ and the 
correlation\footnote{Ref.~\cite{Nadolsky:2008zw} instead uses the
notation $\cos\varphi$ to denote the correlation between two
PDFs/observables.} $\rho\lc \sigma(Z),\sigma(W^{\pm}) \rc$ are given
in Tab.~\ref{tab:ZWratio}, together with the results obtained using the
NNPDF 1.0, CTEQ6.5 and CTEQ 6.6 sets. In the Hessian approach, which is used
for the CTEQ sets, the correlation between the two observables considered,
$\sigma(W^{\pm})$ and $\sigma(Z)$, is computed using the method
described in~\cite{Nadolsky:2008zw}. In the Monte Carlo
approach, used for computing PDF uncertainties for the NNPDF sets, the corresponding expression 
is given, as described in Ref.~\cite{Ball:2008by}, by
\begin{equation}
\label{eq:corr}
  \rho\lc \sigma(Z),\sigma(W^{\pm}) \rc =
  \frac{\la  \sigma(Z)\sigma(W^{\pm})\ra_{\rep}
    -\la  \sigma(Z)\ra_{\rep}\la \sigma(W^{\pm})\ra_{\rep}}{
    \sqrt{\la  \sigma(Z)^2\ra_{\rep} -\la  \sigma(Z)\ra^2_{\rep}}
    \sqrt{\la  \sigma(W^{\pm})^2\ra_{\rep} -\la  \sigma(W^{\pm})\ra^2_{\rep}}}\ ,
\end{equation}
where the averages are performed over the $N_{\rm rep}$ replicas of
the NNPDF sets.

\begin{table}[t]
  \centering
  {\scriptsize
    \begin{tabular}[c]{|c|c|c|c|c|}
      \hline
      & NNPDF1.2           & NNPDF1.0           & CTEQ6.6        & CTEQ6.5         \\
      \hline   
      $r_{ZW}$   & $0.0961 \pm 0.0005$ & $0.0965 \pm 0.0003$ & $0.0964 \pm 0.0004$ & $0.0957 \pm 0.0002$ \\
      \hline
      $\rho\lc \sigma(Z),\sigma(W^{\pm}) \rc$ & $0.976$ & $0.994$ &  $ 0.983$ & $0.994$\\
      \hline
    \end{tabular}
  }
  \caption{\small 
    Comparison of the values for the ratio of the Z and W cross sections
    at the LHC as well as  their correlation $\rho\lc \sigma(Z),\sigma(W^{\pm}) \rc$, 
    Eq.~\ref{eq:corr},  computed with different PDF sets.
    Again, all results are computed for a center-of-mass energy of $\sqrt{s}$=14 TeV.}
  \label{tab:ZWratio}
\end{table}

\begin{figure}[t]
  \centering
  \includegraphics[width=8cm]{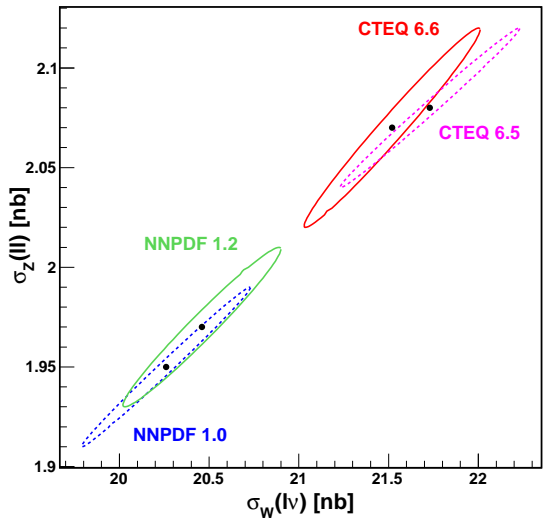}
  \caption{\small Comparison of the $W$ and $Z$ one sigma 
   correlation ellipses at the LHC 
    obtained from different fits: NNPDF 1.2 (green), NNPDF 1.0 (blue), 
    CTEQ 6.6 (red) and CTEQ 6.5 (purple).}
  \label{fig:WZxsec}
\end{figure}

In Fig.~\ref{fig:WZxsec} we compare the $\sigma_Z$-$\sigma_W$  one
sigma correlation ellipses for the NNPDF 1.2, NNPDF 1.0, CTEQ 6.6 and
CTEQ 6.5 sets. We note that, despite the fact that the error band on the
strange parton densities is in general much larger for the NNPDF 1.2
set than for CTEQ 6.6, the uncertainty on the ratio $r_{ZW}$ is of the
same size. 
This is a consequence of the fact that, as previously mentioned, this
ratio is mostly correlated to the strange PDFs in a limited region of
$x$, namely $0.01<x<0.05$, where the NNPDF1.2 and CTEQ6.6 uncertainties on
$s^+$ are  roughly of the same size. This can be
 understood as 
a consequence of the fact that the NuTeV dimuon data, 
which constrains the strangeness 
in the two analyses, cover precisely this kinematical range.

\begin{figure}[t]
  \centering
  \includegraphics[width=0.49\tw]{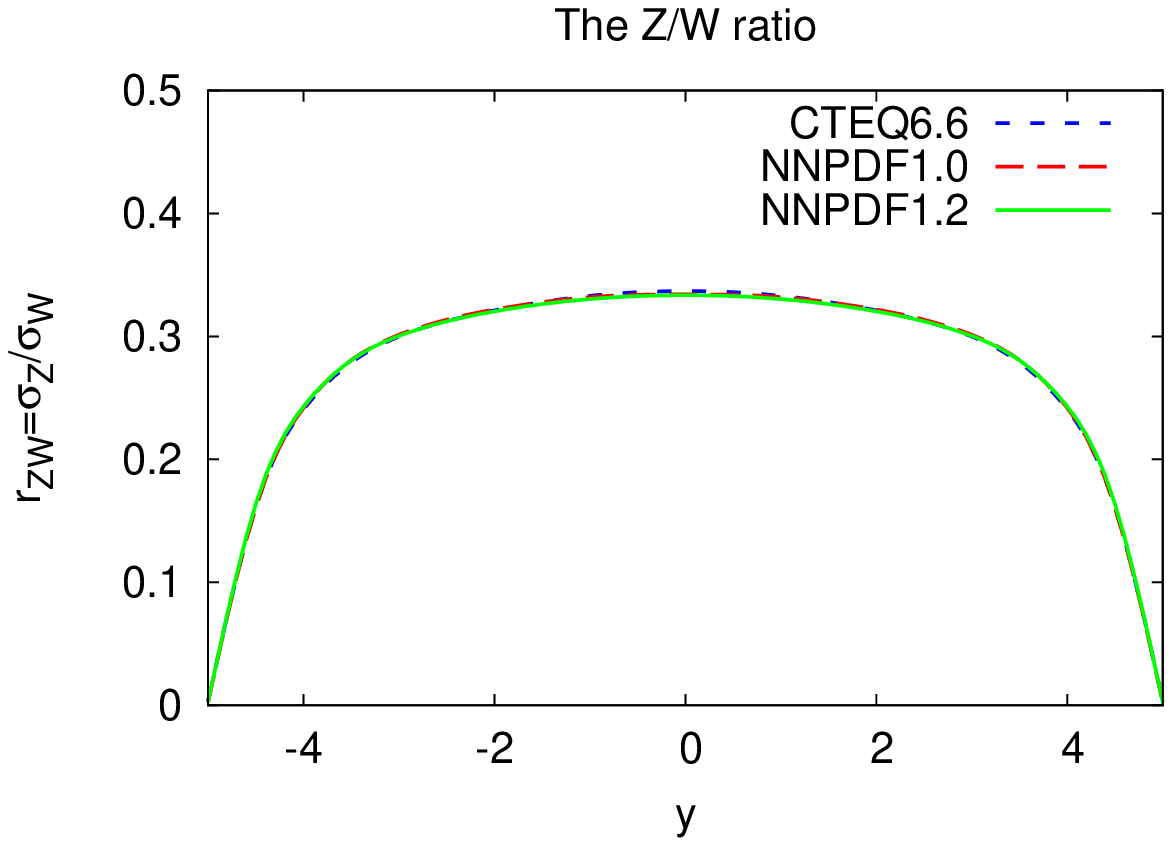}
  \includegraphics[width=0.49\tw]{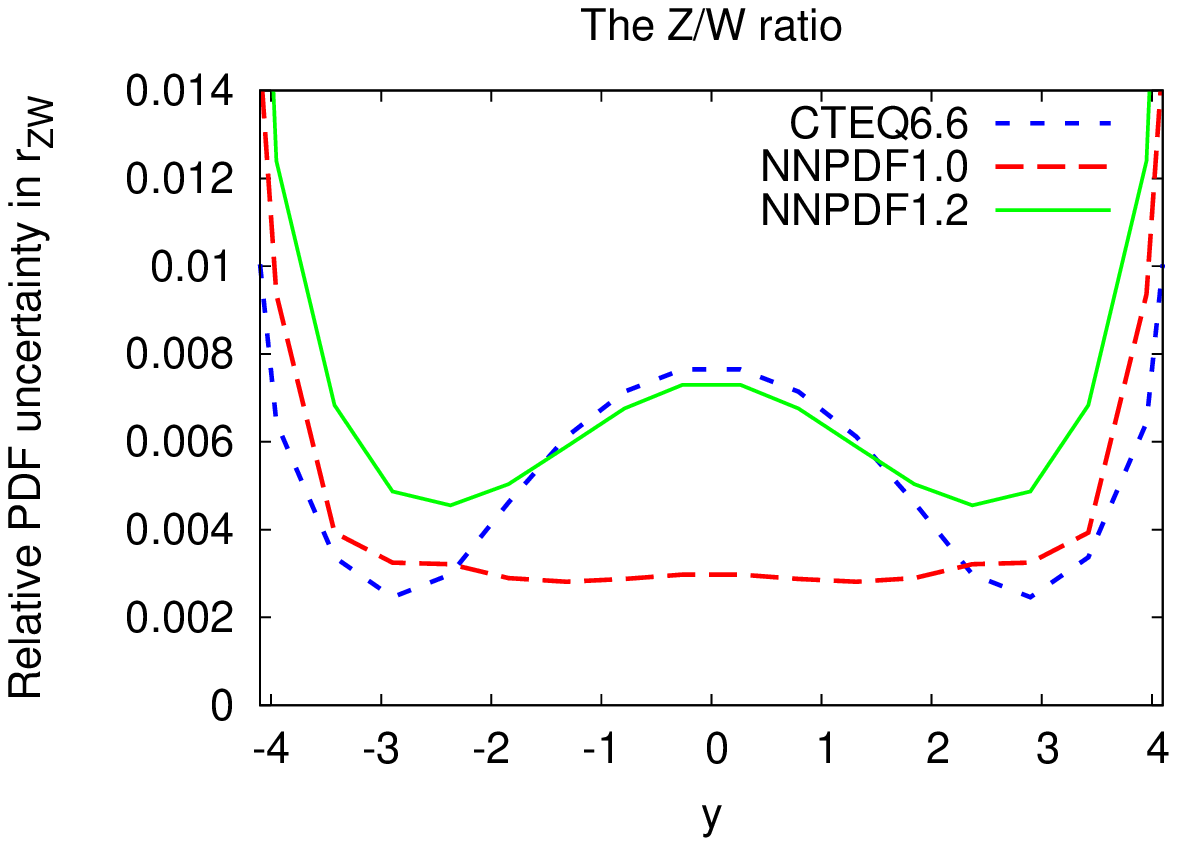}
  \caption{\small Left: the differential rapidity distribution of the
    ZW ratio, defined in Eq.~\ref{eq:wzrat}. The very small PDF
    uncertainties are not shown. Right: the relative PDF uncertainties
    in $r_{ZW}$, as a function of the rapidity $y$.}
  \label{fig:wzrat}
\end{figure}

The situation could be different if we look at the the differential rapidity
distribution.
In order to check this, in Fig.~\ref{fig:wzrat} we show
the rapidity distribution of the ratio  $r_{ZW}$ defined as
\begin{equation}
  \label{eq:wzrat}
  \frac{dr_{ZW}}{dy}\lp y\rp \equiv \frac{d\sigma^Z(y)/dy}{
    d\sigma^W(y)/dy} \ ,
\end{equation}
together with the associated PDF uncertainties.
We observe a sizable increase in the PDF uncertainty for the NNPDF 1.2 
result, when compared to results obtained with other sets, at forward rapidities.
This is due to the increase of the NNPDF1.2 strange
 uncertainties at small-$x$, shown in 
Fig.~\ref{fig:s+}. However in the central rapidity region,
which provides the bulk of the contribution to $r_{ZW}$, the
uncertainties of CTEQ6.6 and NNPDF1.2 turn out to be comparable,
confirming the agreement of PDF uncertainties on the integrated
ratio shown in 
Table~\ref{tab:ZWratio}.

\section{The $t\bar{t}$ total cross section and partonic fluxes}

\begin{figure}[t]
  \centering
  \includegraphics[width=0.70\tw]{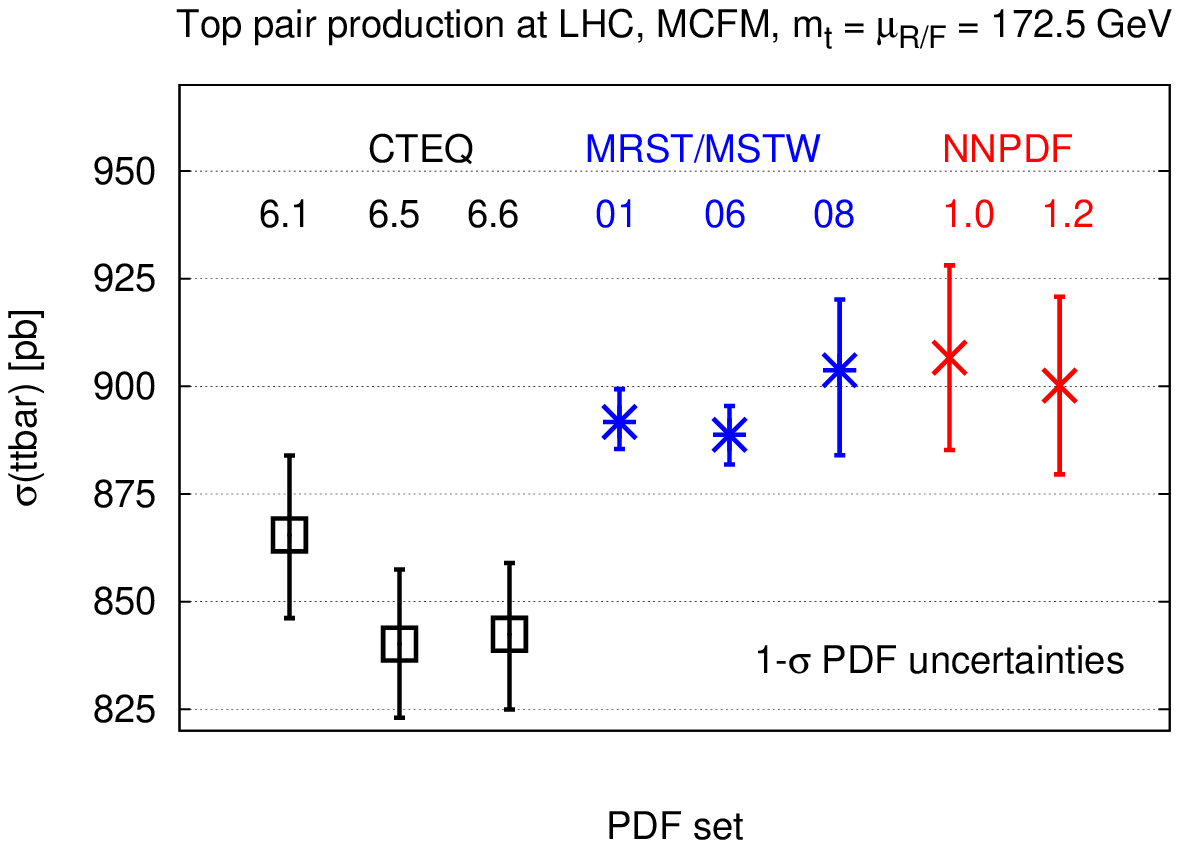}
  \caption{\small Comparison of the total $t\bar{t}$  cross section
    computed with recent PDF sets. The computation
    has been performed with the {\tt MCFM} program at NLO for
    $m_t=\mu_R=\mu_F=172.5$ GeV. All PDF uncertainties
    correspond to 68\% confidence levels.}
  \label{fig:ttbar}
\end{figure}

The total $t\bar{t}$  cross section, $\sigma\lp t\bar{t}\rp$, has been
the subject of several recent 
studies~\cite{Nadolsky:2008zw,Cacciari:2008zb,Kidonakis:2008mu}, 
partially motivated by the
possibility of using this process as a standard candle at the LHC to
measure the luminosity. To compare the predictions
of the NNPDF1.2 set to ones of the other parton determinations,  
the total $t\bar{t}$ cross section is shown in Fig.~\ref{fig:ttbar}
for various PDF sets.

The computation has been performed with MCFM at NLO. No soft
gluon resummation corrections, as done in Ref.~\cite{Cacciari:2008zb}, 
are included.
 All scales are set
equal to $m_t=172.5$ GeV. 
Together with NNPDF1.2, we show the predictions from
NNPDF 1.0 and various other sets from the CTEQ and
MRST/MSTW collaborations. It is clear from  Fig.~\ref{fig:ttbar},
comparing for example NNPDF 1.0 and 1.2, or CTEQ 6.5 and 6.6,
that strangeness plays a rather minor role for this
observable.

In order to understand the results of Fig.~\ref{fig:ttbar}, one
should take into account the fact that the values of $\alpha_s\lp M_Z^2\rp$
used each global PDF are different.
This is especially important for hadronic cross sections in which the leading order process is a strong
interaction process (like in $t\bar{t}$ or $gg\to H$),
since in these cases the cross section at leading order will be proportional to
$\alpha_S^2$. Moreover, in general there is a correlation between $\alpha_s$ and
the PDF uncertainties, mostly the gluon, as discussed for example in Refs.~\cite{Pumplin:2005rh,Martin:2009bu}.
Note that, while CTEQ 6.6 and NNPDF 1.2 use the
value of $\alpha_S(M_Z^2)$ roughly along the lines of the most
up-to-date global average ($\alpha_s=0.118$ and $\alpha_s=0.119$
respectively), MSTW08~\cite{Martin:2009bu} determines it from the 
PDF analysis ending up with $\alpha_S=0.1202^{+0.0012}_{-0.0015}$ at NLO.

The use of different values of $\alpha_S$ has to
be taken into account when comparing this kind
of standard candle processes.
For example, the ratio $\lc \alpha_S^{\rm mstw08}/\alpha_s^{\rm cteq66}\rc^2$,
which appears in the Born top pair production cross-section, implies that, even for 
identical PDFs, the MSTW08 cross-section will
be $\sim 4\%$ higher than the CTEQ6.6 one. Therefore,
the use of different values of $\alpha_S$ could explain
part of the discrepancy between various PDF sets. This emphasizes
the importance of using a unique value of the strong coupling
for comparison between benchmark observables, especially
those proportional to $\alpha_S$ already at leading order.

In order to compare the contributions from each PDF flavour to the
total cross-section for $t\bar{t}$ production we follow the
approach of Ref.~\cite{Nason:1987xz}, 
where the hadronic cross section is written as
\begin{equation}
  {\sigma\lp S,m^2_t\rp} = \frac{\alpha_s^2\lp \mu^2\rp}{m_t^2}
  \sum_{ij}\int_{\rho^t}^1 \frac{d\tau}{\tau}\Phi_{ij}\lp \tau,\mu^2 \rp
  \widetilde{\sigma}_{ij}\lp \frac{\rho^t}{\tau},\frac{\mu^2}{m_t^2}\rp \ ,
  \quad
  \rho^t=\frac{4m_t^2}{S} \ ,
\end{equation}
that is, as a convolution between a partonic cross section
$\widetilde{\sigma}_{ij}$
and parton fluxes $\Phi_{ij}$, defined as
\begin{equation}
  \label{ref:fluxes}
  { \Phi_{ij}\lp \tau,\mu^2 \rp} = \int_0^1 dx_1\int_0^1 dx_2
  { q_i\lp x_1,\mu^2\rp
    q_j\lp x_2,\mu^2\rp}\delta\lp x_1x_2-\tau\rp \ ,
\end{equation}
which in turn are convolutions of parton distributions. Note that
at the LHC, for $\sqrt{S}=14$ TeV,
for a top quark mass of $m_t=172.5$ GeV one obtains
$\rho^t\sim 6\cdot 10^{-4}$.

\begin{figure}[t]
  \centering
  \includegraphics[width=0.49\tw]{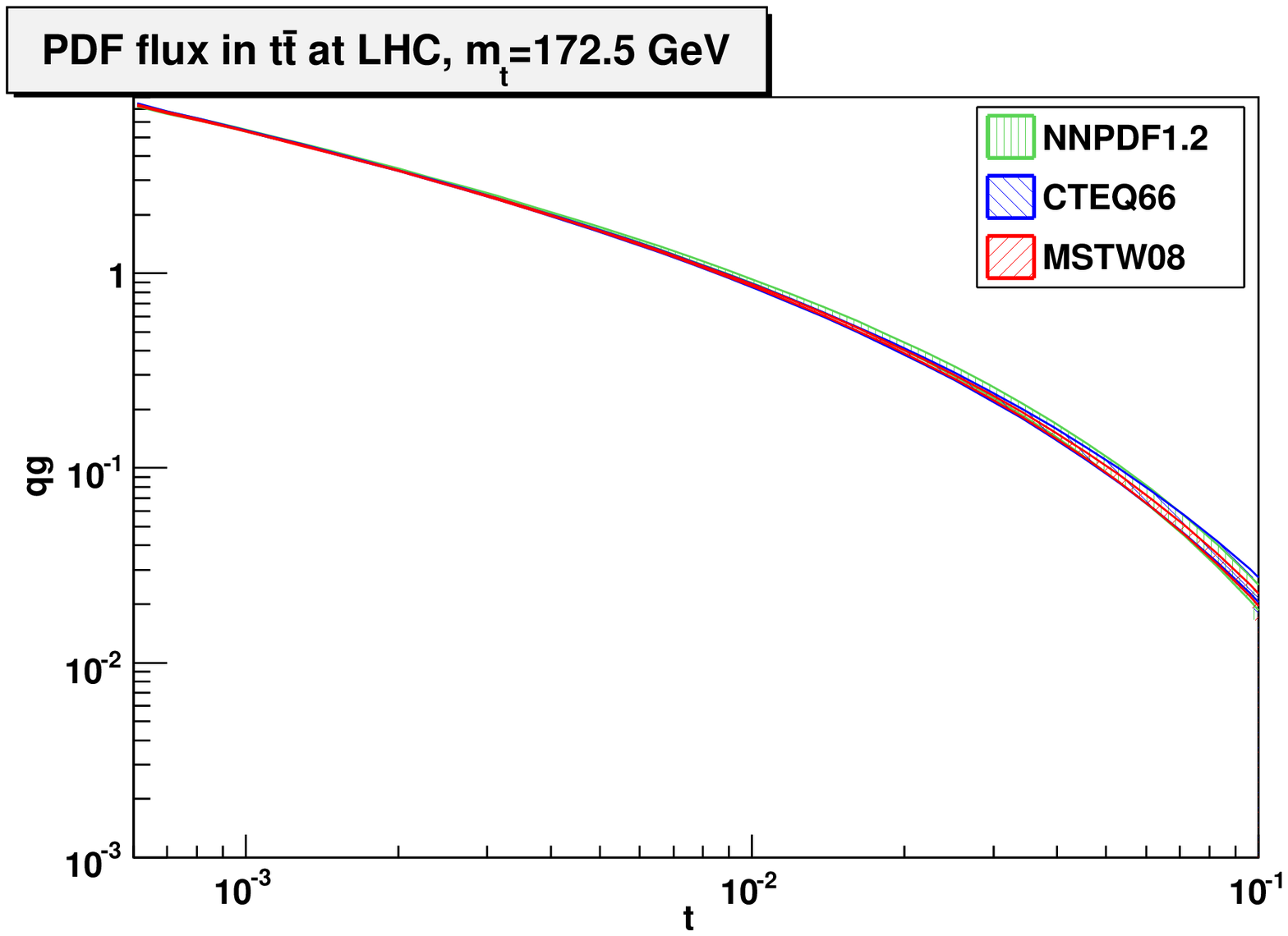}
  \includegraphics[width=0.49\tw]{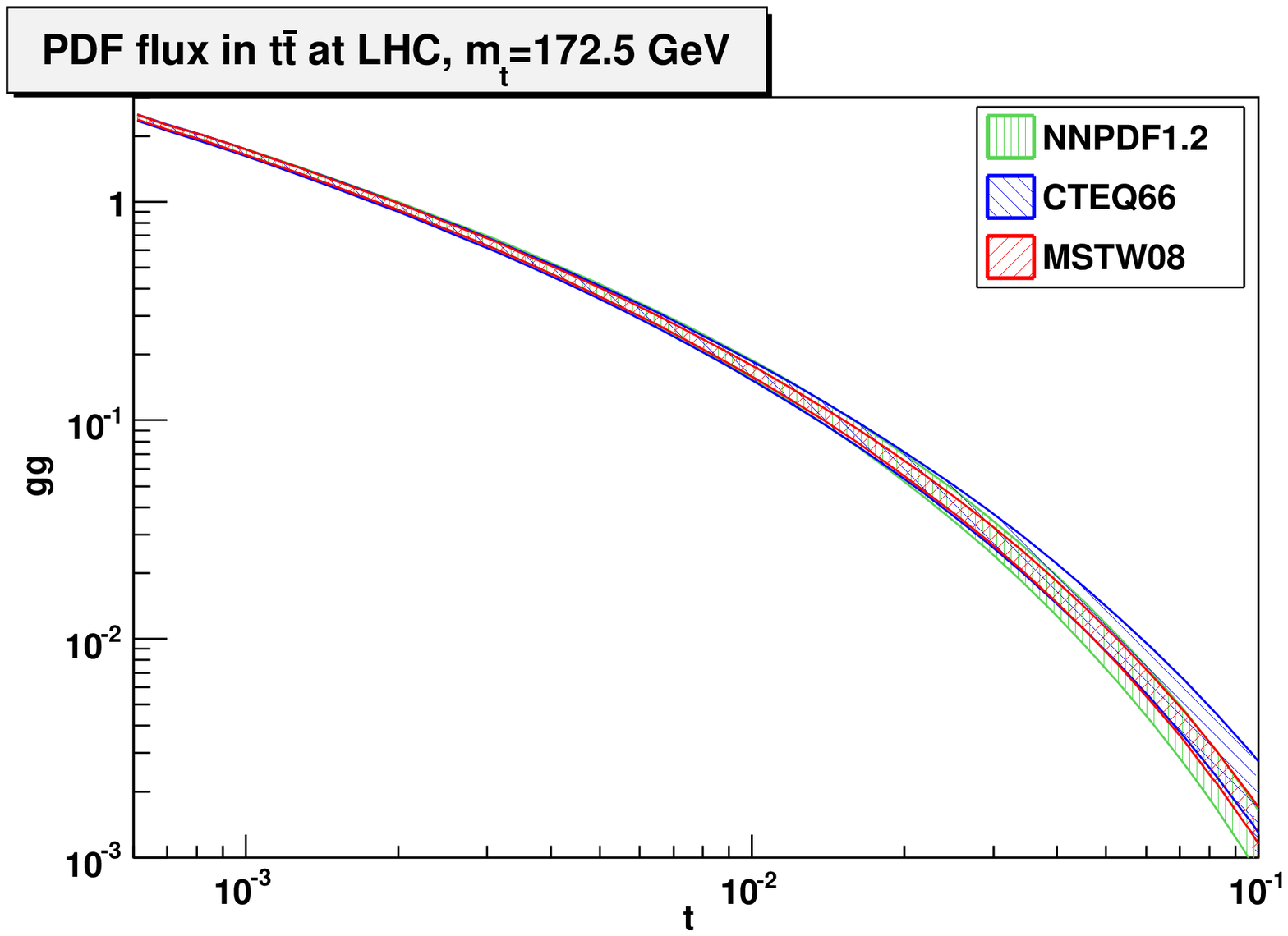}
  \caption{\small The partonic fluxes, defined in Eq.~\ref{ref:fluxes},
in the QG (left) and the GG (right) channel, for the NNPDF1.2,
CTEQ6.6 and MSTW08 PDF sets. The rapid decrease at large $\tau$
reflects the smallness of the PDFs at large $x$. Note that
the partonic fluxes are evaluated at $\mu^2=m_t^2$.}
  \label{fig:fluxes}
\end{figure}

\begin{figure}[t]
  \centering
  \includegraphics[width=0.99\tw]{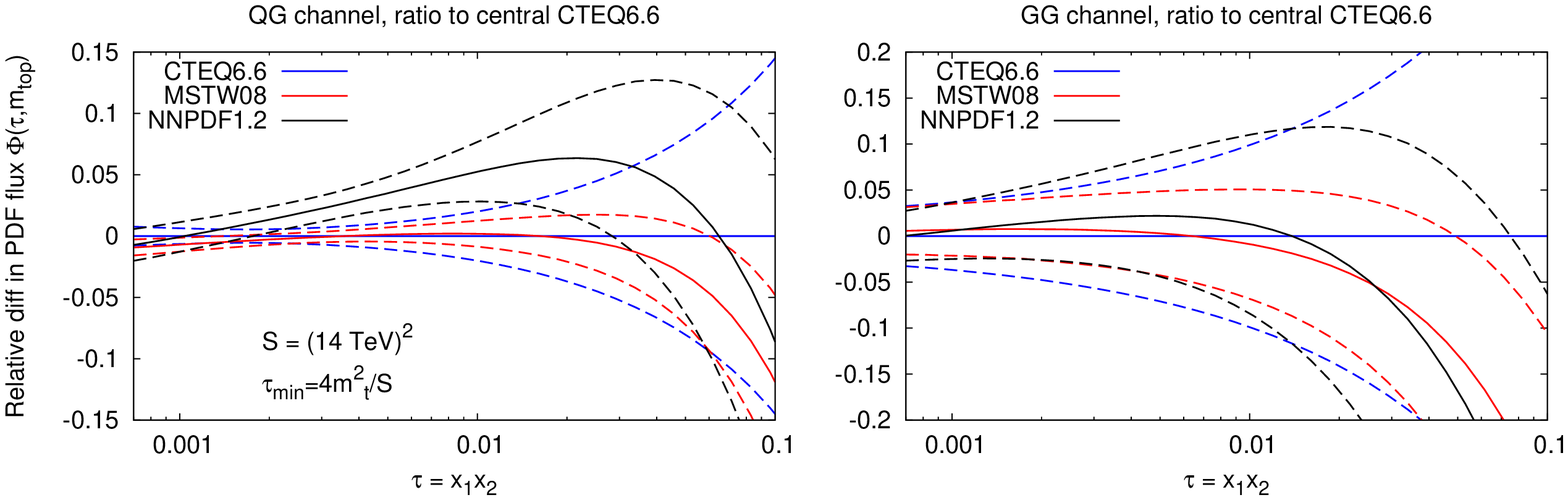}
  \caption{\small The relative differences between the
PDF fluxes in the QG and GG channels between the various
sets considered, in the kinematical region relevant for
top pair production at the LHC. As in Fig.~\ref{fig:fluxes},
 partonic fluxes are evaluated at $\mu^2=m_t^2$. Only the kinematically
allowed region $\tau\ge \tau_{\rm min}=4m_t^2/S$ is shown. }
  \label{fig:fluxes2}
\end{figure}

The dominant contribution to top pair production at the LHC
comes from the gluon-gluon and gluon-quark channels. In Fig.~\ref{fig:fluxes} 
we show the absolute fluxes for these two channels, as defined
in Eq.~\ref{ref:fluxes} for $\mu^2=m_t^2$, for the three most recent PDF
sets of the CTEQ, MSTW~\cite{Martin:2009iq} 
and NNPDF Collaborations, in the kinematical region relevant
for top pair production.
The three PDF sets considered agree reasonably well. In order
to obtain a more detailed picture of the comparison,
in Fig.~\ref{fig:fluxes2} we show the relative differences between
the PDF fluxes for the two channels considered before, using the central
CTEQ6.6 result as a reference. We observe a good agreement for
the fluxes in the GG channel, as well as for the QG channel in most
of the $\tau$ range, with the exception of the middle
region where a 2-$\sigma$ discrepancy is found. The origin
of such discrepancy might be related to differences in the
treatment of heavy quark mass effects in NNPDF1.2 compared
to the other sets.

Finally, let us compare in more detail the
size of PDF uncertainties in the parton fluxes. Following Ref.~\cite{Campbell:2006wx},
we compare the relative PDF uncertainties from the fluxes
computed with  NNPDF1.2, CTEQ6.6 and MSTW08 as a function of
$\hat{S}=x_1x_2S$. That is, we compute PDF uncertainties in
$\Phi_{ij}( \tau=\hat{S}/S,\mu^2 )$ as a function of $\hat{S}$.
These parton fluxes have been computed at
$\mu^2=10^4$ GeV$^2$, the typical scale for processes like
$W,Z$ or $H$ production at the LHC, again
assuming $\sqrt{S}=14$ TeV.

\begin{figure}[t]
  \centering
  \includegraphics[width=0.99\tw]{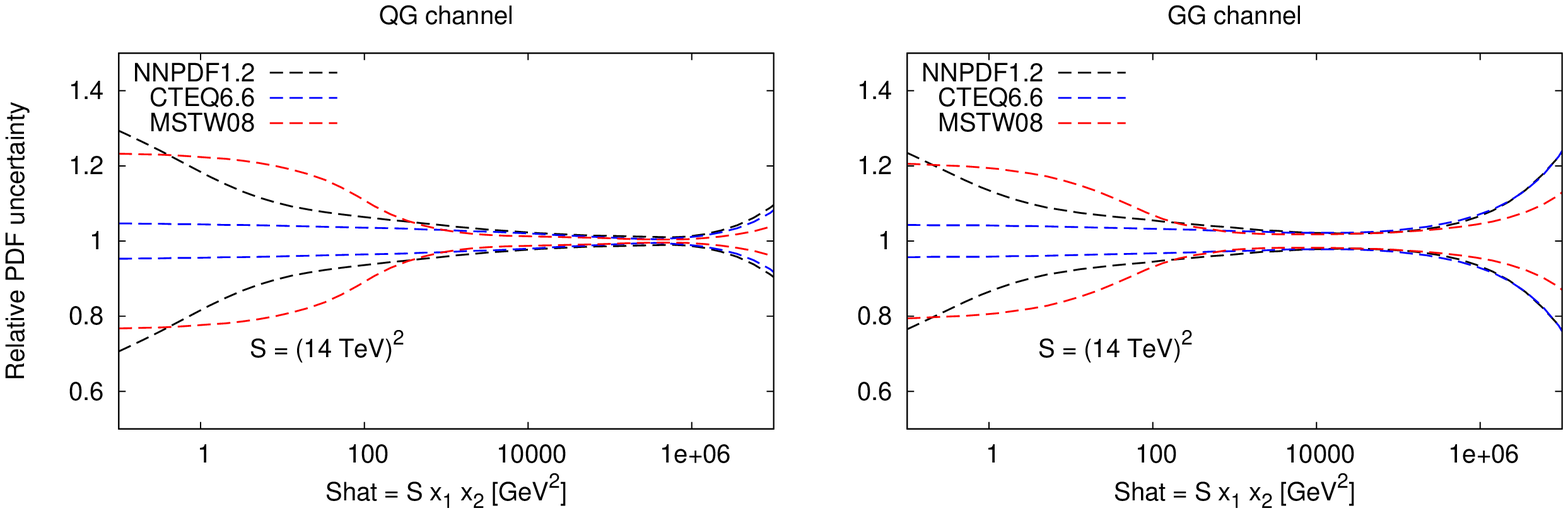}
  \includegraphics[width=0.99\tw]{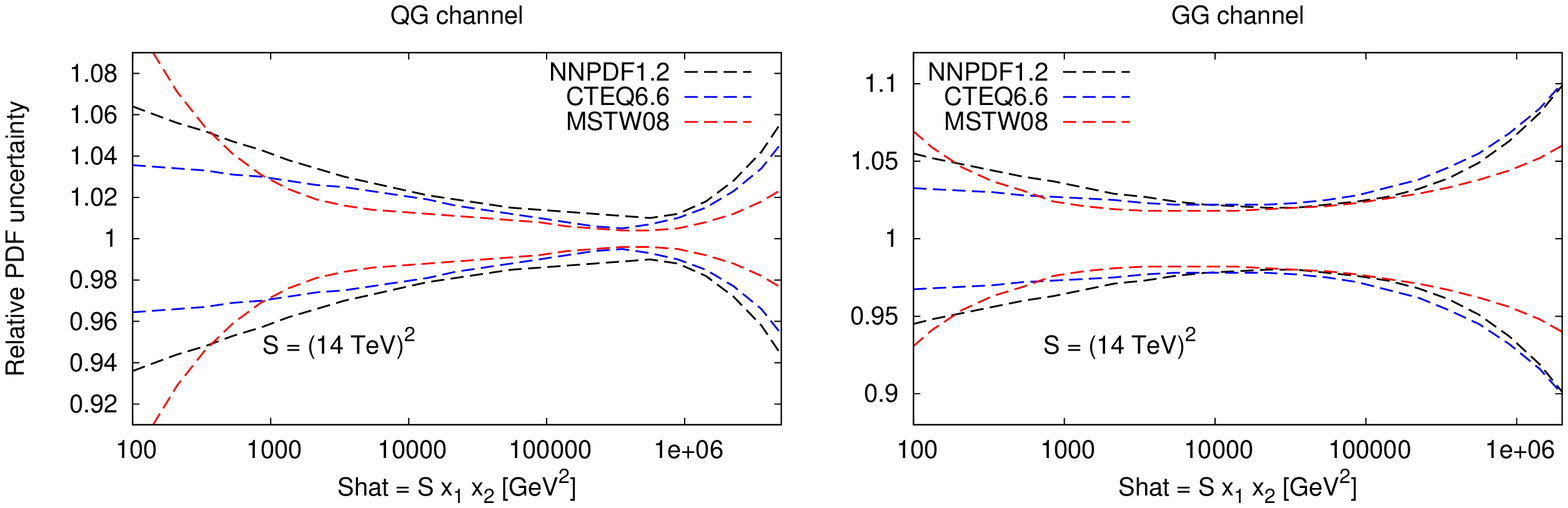}
\caption{\small Comparison between the size of
PDF uncertainties in the partonic fluxes in the
NNPDF1.2, CTEQ6.6 and MSTW08 PDF sets in the region $10^{-5}\le x\le 1$
(above) and a zoom in the middle region (below), where
PDF uncertainties are smallest. Note that in this case
 partonic fluxes are evaluated at $\mu^2=10^4$ GeV$^2$, as
discussed in the text.}
  \label{fig:fluxes3}
\end{figure}

We show the results of the comparison in Fig.~\ref{fig:fluxes3}. 
Note that only PDF
uncertainties are shown, all central values are set to 1,
unlike the case of Fig.~\ref{fig:fluxes}.
We observe a reasonable agreement in the
size of PDF uncertainties in the intermediate $\hat{S}$
region, and sizable differences at smaller $\hat{S}$.
In the intermediate $\hat{S}$ region, relevant for the production
of massive objects,
the PDF uncertainty in the QG channel turns out to 
be somewhat larger in NNPDF1.2 as compared to both
CTEQ6.6 and MSTW08. This could be due to missing hadronic
data in the former or to a parametrization bias in the latter: the
upcoming NNPDF2.0 global analysis should settle this issue.

\section{$Wc$ production and constraining strangeness at the
LHC}

In the last section of this contribution, we focus on a LHC process
which in principle could be used to measure the strange PDF which is presently
poorly constrained in the region $x<10^{-2}$~\cite{nnpdf12},
$Wc$ associated production. 
The associated production of a vector boson and a charm quark at 
hadronic colliders is directly sensitive to the strange sea PDF and for
this reason, in the past it has been proposed as a candidate for
constraining the strange PDF at the TeVatron and the LHC
colliders~\cite{Lai:2007dq}.

We revisit this proposal by comparing the NNPDF1.2 predictions for the total
$Wc$ cross section with recent measurements at the Tevatron~\cite{Aaltonen:2007dm} 
and giving predictions for the LHC. 

The dominant production channel for $Wc$ production is $gq\to Wc$, with $q$ 
a down-type quark. 
As in the case of neutrino dimuon production, the down and bottom 
initiated contributions are suppressed with respect to the 
strange one by the smallness of the corresponding CKM matrix elements,
and therefore at LO, neglecting CKM mixing, the cross section is
proportional to the strange PDF.

Associated $W$ and charm production looks therefore like a
promising channel for providing a direct constraint on the strange PDF at
the energy scale of $M_W$, two orders of magnitude above the typical
energy of the NuTeV dimuon data. 

This process has already been studied at the Tevatron.
In particular, the CDF  experiment has published a measurement of the $Wc$ production 
cross section, obtained using $\sim 1.8$ 
fb$^{-1}$ of $p\overline{p}$ collisions at $\sqrt{s}=1.96$ TeV.
A NLO prediction for this observable can be obtained easily using the MCFM~\cite{MCFMurl} 
code. 

The result obtained by CDF for the  measurement of the total cross section is 
\begin{equation}
  \label{eq:Wc-CDF-result}
  \sigma_{Wc}(p_{Tc}>20\mathrm{GeV},|\eta_c|<1.5)\times 
  \mathrm{BR} (W\to l\nu) =
  9.8 \pm 3.2 \mathrm{pb},\,
\end{equation}
which is in agreement with the NLO prediction obtained using MCFM and the 
NNPDF 1.2 set:
\begin{equation}
  \label{eq:Wc-NNPDF-result}
  \sigma_{Wc}(p_{Tc}>20\mathrm{GeV},|\eta_c|<1.5)
  \times \mathrm{BR} (W\to l\nu) =
  10.11 \pm 1.24 (\mathrm{PDF}) 
  {}^{+0.74}_{-0.92}(\mathrm{scale})\;
  \mathrm{pb},
\end{equation}
where the first error is the one coming from PDF uncertainties and the second one is
due to the variation of the renormalization and factorization scales 
in the perturbative computation. 
The expected precision of the experimental result, when extrapolated to the full
Run II dataset ($\sim 6-7$ fb$^{-1}$), is $\sim 15\%$, comparable to the 
present uncertainty on the theoretical prediction. The
theoretical uncertainties at the Tevatron is dominated by
PDF uncertainties, with a sizable contribution from scale
dependence.

In order to investigate the possibility of using this very same
channel as a strangeness constraint at the LHC, 
we also computed the $Wc$ cross section for the LHC assuming a 
center of mass energy of 14 TeV and standard rapidity and transverse momentum cuts 
both for the charm quark and the leptons coming from the $W$ decay. The result
that we obtain is
\begin{equation}
  \label{eq:Wc-NNPDF-LHC-result}
  \sigma_{Wc}(p_{Tc}>20\mathrm{GeV}/c,|\eta_c|<4.)
  \times \mathrm{BR} (W\to l\nu) =
  631 \pm 46 (\mathrm{PDF}) {}^{+38}_{-63}(\mathrm{scale})\;
  \mathrm{pb}\,.
\end{equation}
Our result shows that the uncertainty on the theoretical prediction due to 
scale variations, and thus to higher order corrections, is
comparable to the uncertainty due to the strange PDFs. 
The result seems to suggest that
it might be difficult to use this process to constrain PDFs. 
It might instead be useful to look at differential distributions, 
like the $W$ or $c$-jet rapidity or transverse momentum distributions. In any case, unless
scale uncertainties can be reduced by higher order
computations, the theoretical error limits the usefulness
of $Wc$ production as a constraint for the strangeness distributions,
regardless the accuracy of present and
future experimental measurements of this channel.

\section{Outlook}

In this contribution we reviewed some phenomenological implications of the 
NNPDF1.2 parton set for LHC physics, with emphasis on those
observables more directly sensitive to the strange PDFs. A more detailed
study of the phenomenological implications of the NNPDF
approach to LHC observables is however postponed until
the release of the upcoming NNPDF2.0 global parton analysis, which
includes all relevant hadronic data like Drell-Yan pair production,
vector boson production and inclusive jet production.

\paragraph{Acknowledgments}

We would like to thank A. Vicini and G. Ridolfi for providing us their
code for the computation of the Drell-Yan process and M. Cacciari
for useful comments.
J.R. thanks the hospitality of the CERN TH division
where part of this work was completed. M.~U. is founded by a SUPA graduate studentship.

\begin{footnotesize}

\providecommand{\href}[2]{#2}\begingroup\raggedright\endgroup

\end{footnotesize}


\begin{thebibliography}{10}

\bibitem{f2ns}
S.~Forte, L.~Garrido, J.~I. Latorre, and A.~Piccione, ``Neural network
  parametrization of deep-inelastic structure functions,'' {\em JHEP} {\bf 05}
  (2002)  062,
\href{http://arxiv.org/abs/hep-ph/0204232}{{\tt hep-ph/0204232}}.

\bibitem{f2p}
{\bf NNPDF} Collaboration, L.~Del~Debbio, S.~Forte, J.~I. Latorre, A.~Piccione,
  and J.~Rojo, ``Unbiased determination of the proton structure function f2(p)
  with faithful uncertainty estimation,'' {\em JHEP} {\bf 03} (2005)  080,
\href{http://arxiv.org/abs/hep-ph/0501067}{{\tt hep-ph/0501067}}.

\bibitem{DelDebbio:2007ee}
{\bf NNPDF} Collaboration, L.~Del~Debbio, S.~Forte, J.~I. Latorre, A.~Piccione,
  and J.~Rojo, ``{Neural network determination of parton distributions: The
  nonsinglet case},'' {\em JHEP} {\bf 03} (2007)  039,
\href{http://arxiv.org/abs/hep-ph/0701127}{{\tt arXiv:hep-ph/0701127}}.

\bibitem{Ball:2008by}
{\bf NNPDF} Collaboration, R.~D. Ball {\em et al.}, ``{A determination of
  parton distributions with faithful uncertainty estimation},''
  \href{http://dx.doi.org/10.1016/j.nuclphysb.2008.09.037}{{\em Nucl. Phys.}
  {\bf B809} (2009)  1--63},
\href{http://arxiv.org/abs/0808.1231}{{\tt arXiv:0808.1231 [hep-ph]}}.

\bibitem{Rojo:2008ke}
{\bf NNPDF} Collaboration, J.~Rojo {\em et al.}, ``{Update on Neural Network
  Parton Distributions: NNPDF1.1},''
\href{http://arxiv.org/abs/0811.2288}{{\tt arXiv:0811.2288 [hep-ph]}}.

\bibitem{nnpdf12}
{\bf NNPDF} Collaboration, R.~D. Ball {\em et al.}, ``{Precision determination
  of electroweak parameters and the strange content of the proton from neutrino
  deep-inelastic scattering},''
\href{http://arxiv.org/abs/0906.1958}{{\tt arXiv:0906.1958 [hep-ph]}}.

\bibitem{Dittmar:2009ii}
M.~Dittmar {\em et al.}, ``{Parton Distributions},''
\href{http://arxiv.org/abs/0901.2504}{{\tt arXiv:0901.2504 [hep-ph]}}.

\bibitem{Nadolsky:2008zw}
P.~M. Nadolsky {\em et al.}, ``{Implications of CTEQ global analysis for
  collider observables},''
  \href{http://dx.doi.org/10.1103/PhysRevD.78.013004}{{\em Phys. Rev.} {\bf
  D78} (2008)  013004},
\href{http://arxiv.org/abs/0802.0007}{{\tt arXiv:0802.0007 [hep-ph]}}.

\bibitem{Tung:2006tb}
W.~K. Tung {\em et al.}, ``{Heavy quark mass effects in deep inelastic
  scattering and global QCD analysis},'' {\em JHEP} {\bf 02} (2007)  053,
\href{http://arxiv.org/abs/hep-ph/0611254}{{\tt arXiv:hep-ph/0611254}}.

\bibitem{Pumplin:2002vw}
J.~Pumplin {\em et al.}, ``{New generation of parton distributions with
  uncertainties from global QCD analysis},'' {\em JHEP} {\bf 07} (2002)  012,
\href{http://arxiv.org/abs/hep-ph/0201195}{{\tt arXiv:hep-ph/0201195}}.

\bibitem{Cacciari:2008zb}
M.~Cacciari, S.~Frixione, M.~L. Mangano, P.~Nason, and G.~Ridolfi, ``{Updated
  predictions for the total production cross sections of top and of heavier
  quark pairs at the Tevatron and at the LHC},''
  \href{http://dx.doi.org/10.1088/1126-6708/2008/09/127}{{\em JHEP} {\bf 09}
  (2008)  127},
\href{http://arxiv.org/abs/0804.2800}{{\tt arXiv:0804.2800 [hep-ph]}}.

\bibitem{Kidonakis:2008mu}
N.~Kidonakis and R.~Vogt, ``{The Theoretical top quark cross section at the
  Tevatron and the LHC},''
  \href{http://dx.doi.org/10.1103/PhysRevD.78.074005}{{\em Phys. Rev.} {\bf
  D78} (2008)  074005},
\href{http://arxiv.org/abs/0805.3844}{{\tt arXiv:0805.3844 [hep-ph]}}.

\bibitem{Pumplin:2005rh}
J.~Pumplin, A.~Belyaev, J.~Huston, D.~Stump, and W.~K. Tung, ``{Parton
  distributions and the strong coupling: CTEQ6AB PDFs},'' {\em JHEP} {\bf 02}
  (2006)  032,
\href{http://arxiv.org/abs/hep-ph/0512167}{{\tt arXiv:hep-ph/0512167}}.

\bibitem{Martin:2009bu}
A.~D. Martin, W.~J. Stirling, R.~S. Thorne, and G.~Watt, ``{Uncertainties on
  $\alpha_S$ in global PDF analyses},''
\href{http://arxiv.org/abs/0905.3531}{{\tt arXiv:0905.3531 [hep-ph]}}.

\bibitem{Nason:1987xz}
P.~Nason, S.~Dawson, and R.~K. Ellis, ``{The Total Cross-Section for the
  Production of Heavy Quarks in Hadronic Collisions},''
\href{http://dx.doi.org/10.1016/0550-3213(88)90422-1}{{\em Nucl. Phys.} {\bf
  B303} (1988)  607}.

\bibitem{Martin:2009iq}
A.~D. Martin, W.~J. Stirling, R.~S. Thorne, and G.~Watt, ``{Parton
  distributions for the LHC},''
\href{http://arxiv.org/abs/0901.0002}{{\tt arXiv:0901.0002 [hep-ph]}}.

\bibitem{Campbell:2006wx}
J.~M. Campbell, J.~W. Huston, and W.~J. Stirling, ``{Hard interactions of
  quarks and gluons: A primer for LHC physics},''
  \href{http://dx.doi.org/10.1088/0034-4885/70/1/R02}{{\em Rept. Prog. Phys.}
  {\bf 70} (2007)  89},
\href{http://arxiv.org/abs/hep-ph/0611148}{{\tt arXiv:hep-ph/0611148}}.

\bibitem{Lai:2007dq}
H.~L. Lai {\em et al.}, ``{The Strange Parton Distribution of the Nucleon:
  Global Analysis and Applications},'' {\em JHEP} {\bf 04} (2007)  089,
\href{http://arxiv.org/abs/hep-ph/0702268}{{\tt arXiv:hep-ph/0702268}}.

\bibitem{Aaltonen:2007dm}
{\bf CDF} Collaboration, T.~Aaltonen {\em et al.}, ``{First Measurement of the
  Production of a W Boson in Association with a Single Charm Quark in Proton
  Anti-proton Collisions at sqrt(s)=1.96 TeV},''
  \href{http://dx.doi.org/10.1103/PhysRevLett.100.091803}{{\em Phys. Rev.
  Lett.} {\bf 100} (2008)  091803},
\href{http://arxiv.org/abs/0711.2901}{{\tt arXiv:0711.2901 [hep-ex]}}.

\bibitem{MCFMurl}
{\em MCFM}.
\newblock \href{http://arxiv.org/abs/http://mcfm.fnal.gov}{{\tt
  http://mcfm.fnal.gov}}.
\newblock \url{http://mcfm.fnal.gov}.

\end{thebibliography}
\end{document}